\documentclass[%
reprint,
nofootinbib,
amsmath,amssymb,
prd,
]{revtex4-1}
\usepackage{datetime}
\usepackage{natbib}
\usepackage{hyperref}
\usepackage{color}
\usepackage{ulem}

\newcommand{\mc}{\mathcal}
\newcommand{\f}[2]{\frac{#1}{#2}}

\begin{document}

\title{Comment on ``Boosted Kerr black holes in general relativity''}

\author{Emanuel Gallo$^{1,2}$}
\email{egallo@unc.edu.ar}
\affiliation{$^1$FaMAF, UNC; $^2$Instituto de F\'isica Enrique Gaviola (IFEG), CONICET, \\
Ciudad Universitaria, (5000) C\'ordoba, Argentina. }

\author{Thomas M\"adler$^{3}$}
\email{thomas.maedler@mail.udp.cl}\affiliation{$^3$Escuela de Obras Civiles and N\'ucleo de Astronom\'ia, Facultad de Ingenier\'{i}a y Ciencias, Universidad Diego Portales, Avenida Ej\'{e}rcito
Libertador 441, Casilla 298-V, Santiago, Chile.}

\date{\today}

\begin{abstract}
We discuss a recently presented boosted Kerr black hole solution which  had already been used by other authors. This boosted metric is based on wrong assumptions regarding asymptotic inertial observers and moreover the performed boost is not a proper Lorentz transformation. 
This note aims to clarify some of the issues when boosting black holes and the necessary care in order to interpret them.
As it is wrongly claimed that the presented boosted Kerr metric is of Bondi-Sachs type, we recall out some of the necessary requirements and difficulties, when the  casting the Kerr metric into a metric with a surface forming null coordinate.
\end{abstract}

\pacs{Valid PACS appear here}

\maketitle

\section{Introduction}

Boosted black holes are relevant in gravitational physics. For example, the final black hole remnant of a binary black hole merger is in general boosted with respect to the rest frame of the two initial black holes. 
This property has important bearing for gravitational wave physics as is gives rise for an additional observable in gravitational wave astronomy --  the gravitational wave memory \cite{zeldovic,brag,bontz, thorne,christ}, which is the permanent displacement of test masses after the passage of a gravitational wave. 
This memory effect can be decomposed into a two parts -  an ordinary or linear memory effect related to a boost \cite{skypattern,ST_RM&AM} and a null memory effect related to the loss of energy of the radiating system by massless particles (electromagnetic radiation \cite{bieri_em,tolish}, neutrinos \cite{bieri_nu} or gravitons \cite{thorne})\footnote{The list of references on gravitational wave memory physics is by far  not complete, since this is not a review note on gravitational wave memory. We
apologize for our arbitrary choice of references.}.
In particular, the extraction of physical observables like the gravitational wave memory as well as the ``classical'' observables like gravitational radiation \cite{Bondi,BSScholar}, linear and angular momentum \cite{momentum,waveform,AM_comment} at null infinity needs to be done in a generalization of an inertial frame.
These frames at null infinity are tied to a particular null tetrad and  called Bondi frames. 
The corresponding coordinates are the Bondi coordinates. 
Bondi frames are in general related to one another by  transformations of the Bondi-Metzner-Sachs (BMS) group \cite{SachsSym}, which  include the infinite dimensional subgroup of  supertranslations.
These supertranslations relate different cross sections (``cuts'') of null infinity with each other.
Their existence prevents to single out a  canonical Poincare sub-group at null infinity.  
However for stationary metrics, like the Kerr metric, there exists a canonical way to set a preferred Poincare subgroup based in the notion of good cuts\cite{good-cuts} or its generalization through nice sections 
\cite{Moreschi_1988}.

Since a boost in Special Relativity is done with respect to observers in  inertial frames, it is  clear that an asymptotic  boost in an asymptotically flat spacetime ought be done a with respect to an associated Bondi frames.
Notably, an expression for the Kerr metric approaching a Bondi frame is not known in an explicit closed analytic form. 
One of the reason is that the  principal null directions of the Kerr solution are twisting.
Meaning they do not generate null surfaces. 
Therefore, it is not a simple task to construct a Bondi-like coordinate system. For the asymptotic analysis, a way  to approach a Bondi frame for the Kerr metric at null infinity was archieved in  \cite{BKS} by introduction of a set of hyperboloidal coordinates. 
These coordinates are defined with respect to hypersurfaces that are null at null infinity and spacelike in its neighborhood. 

Recently, an algorithm to construct boosted Kerr  black hole solutions was presented in the peer-reviewed references \cite{soares1,soares2}.
In the first work \cite{soares1}, the author presents a simplified analysis, where the Kerr black hole is boosted along $z-$axis, only. 
The subsequent article \cite{soares2} covers the general boost in arbitrary directions. 
In both situations, the author claims that these solutions represent  boosted Kerr metrics as ``seen" by an asymptotic inertial observer.
The proposed mechanism seems to be simple. 
Thus, making it favorable to use, if physical effects of moving rotating black holes ought  to be  studied. 
Indeed, follow up work of other authors \cite{emBoostSoares,LensingBoostSoares} using these metrics seems to validate them. 

We analyze the metric presented in  \cite{soares1} in greater detail  and clarify some of the issues arising from a misunderstanding of the meaning of an asymptotic inertial observer. 
As the mechanism for the boost in \cite{soares2} uses the same (but more sophisticated) techniques, the faulty assumptions are taken over from \cite{soares1} to \cite{soares2}.
Therefore, the  main  results of \cite{soares2} can be questioned from the same grounds.
We will further show that for the metric presented in  \cite{soares1} (and consequently also for the proposed extension in \cite{soares2}),  it can not be deduced that it is  the coordinate representation of a boosted Kerr metric with respect to an asymptotic Lorentzian observer.
In particular, the discussed metrics  contain an incomplete piece of a Lorentz transformation in a certain sense. 
More precisely, the coordinate representation of the `boosted Kerr metrics' in \cite{soares1,soares2} only make use of an angular coordinate transformation of the original Kerr metric that could be thought as associated to an asymptotic Lorentzian observer.
However, the additional transformations of the timelike and radial coordinates are yet missing. 
Therefore, the chosen coordinates  do not represent adapted coordinates with respect to an inertial observer.
Consequently, care  must be taken in the interpretation of the `boosted' Kerr metrics of \cite{soares1,soares2}, because without the necessary care it can give rise to wrong results with respect to the physics related to moving black holes as measured by asymptotic inertial observers. 
For example, physical effects of a boosted rotating  Kerr black hole (with respect to the proper asymptotic observer) do not differ at leading order from those of a boosted Schwarzschild black hole. 
This is clear, because for large values of the (proper) radial coordinate $r$, the effects of the spin of the black hole enter at higher order of a $1/r^n$ expansion than those resulting from the mass. 
There exist several ways to present a boosted Schwarzschild black hole in the literature. 
Some (e.g. \cite{Aichelburg:1970dh})  use properly adapted coordinates to asymptotic inertial observers, while other make usage of non-inertial coordinates, as for example in terms of Newman-Unti coordinates, which in general do not conform an inertial (Bondi) frame \cite{Dain:1996phot-rocket}. 
In the last case, extra work and significant machinery is needed in order to extract physical information (see e.g. \cite{waveform} or \cite{deadman}).
Another effect that cannot be reproduced by \cite{soares1,soares2} in a straightforward way is the fact that the comparison of an un-boosted Kerr black hole in  its distant past with its boosted  version of it in its distant future gives rise a gravitational wave memory and a corresponding supertranslation \cite{ST_RM&AM}.

\section{Faulty points in the boosted solution}
Here we point out the inconsistencies in \cite{soares1,soares2} that do not capture the physics of asymptotic Lorentz transformation. 
In particular, we show that the mentioned solution can be easily obtained from a simple coordinate transformation in the angular directions applied to the a original Kerr metric. 

With respect to coordinates $\tilde x^\alpha =(\tilde u , \tilde r, \tilde \theta, \tilde \phi)  $,  the outgoing Eddington-Finkelstein form of the Kerr metric is given by  \cite{kerr}\footnote{Note, here are some corrections to the original form in \cite{kerr}. The corrections are pointed out by Kerr himself in \cite{KerrDiscovery}. In particular, the  positive sense of rotation is used. 
Moreover, in Kerr's  original paper the advanced(!) time  is called $u$ \cite{teukolsky}. Kerr's original paper, should be corrected using $u\rightarrow -u$ and $a\rightarrow -a$. } 
\begin{eqnarray}\label{eq:kerr-metric}
    ds^2 &=&
    \Big(\tilde{r}^2+a^2\cos^2\tilde{\theta}\Big)
    \Big(d\tilde\theta^2+\sin^2\tilde\theta d\tilde\phi^2\Big)
    \nonumber\\& 
    -&2\Big(d\tilde u+a\sin^2\tilde\theta d\tilde \phi\Big)
    \Big(d\tilde r-a\sin^2\tilde\theta d\tilde \phi\Big)\nonumber\\
    &
    -&\Big(1-\f{2m\tilde r}{\tilde r^2+a^2\cos^2\tilde\theta}\Big)\Big(d \tilde u+a\sin^2\tilde\theta d\tilde\phi\Big)^2,\\
    &&\nonumber
\end{eqnarray}
where $m$ is the mass and $a$ the specific angular momentum. 
In \cite{soares2}, the most general  `boosted' version of this metric with respect to  coordinates $(u,r,\theta,\phi)$ is presented as  (eq. (27) in \cite{soares2})\footnote{Note, some slight change in notation to be in tune with standard notation for the Kerr metric; to obtain \eqref{eq:boosted_general_Kerr_Soares} in \cite{soares2} make the following substitutions: $a\rightarrow \omega$, $A\rightarrow a$, $B\rightarrow b$.}
\begin{widetext}
\begin{equation}\label{eq:boosted_general_Kerr_Soares}
\begin{aligned}
ds^2=& \frac{r^2 + \Sigma^2}{\mc{K}^2} \Big(d\theta^2+\sin^2\theta d\phi^2\Big)
+\Big(\frac{r^2-2mr +\Sigma^2}{r^2+\Sigma^2}\Big)\Big[d u-2\mc{L}\cot\Big(\f{\theta}{2}\Big)d \phi\Big]^2
\\
-&2\Big[d u-2\mc{L}\cot\Big(\f{\theta}{2}\Big)d \phi\Big]\Big\{dr +\f{a}{\mc{K}^2}[-n_1\sin^2\theta +(n_2\cos\phi+n_3\sin\phi)\sin\theta\cos\theta ]d\phi 
+\f{a}{\mc{K}^2}(n_2\sin\phi-n_3\cos\phi)d\theta\Big\}
    \\
\end{aligned}
\end{equation}
\end{widetext}
where 
\begin{eqnarray}
\mc{K}&=& A+B(\hat x^in_i)\;, A^2-B^2 = 1\\
\Sigma &=& a\frac{B+A(\hat x^in_i)}{A+B(\hat x^in_i)}\\
\mc{L}&=&\Big(\f{1-\cos\theta}{\sin\theta}\Big)\Big(\f{a}{2B^2}-\int\f{\Sigma}{\mc{K}}\sin\theta d\theta\Big),
\end{eqnarray}
with the general direction of the boost $n_i = (n_1,n_2,n_3)$ that is subject to $\delta_{ij} n^in^j=1$, the rapidity $\zeta$ to determine $A = \cosh\zeta$ and $B=\sinh\zeta$, and $\hat x^i = ( \cos\theta,\sin\theta\cos\phi, \sin\theta\sin\phi)$. 
In \cite[page 4]{soares2} it is claimed that {\it ``For $n_2=0=n_3$ and $B=0$ the metric(27)} \footnote{Our Eq. \eqref{eq:boosted_general_Kerr_Soares}.}{\it is the original Kerr metric in retarded Bondi--Sachs--type coordinates.''} In addition, in \cite{[page 1], soares2} is also claimed that {\it ``The derivation
and interpretation of this solution will be framed in the
Bondi-Sachs (BS) characteristic formulation of gravitational wave emission in general relativity, where
we have a clear and complete derivation of physical quantities and its conservation laws...''}. 
Both statements are  not true: regarding the former, an expression for the Kerr metric in  explicit closed form in Bondi-Sachs-type coordinates is not known. 
Concerning the latter, a retarded Bondi coordinate system  is characterized by a surface forming null coordinate  $\hat{u}$ such that  null hypersurfaces $\hat{u}=\text{const}$  are generated by a null geodesic congruence  $\ell_\mu=(d\hat{u})_\mu$ reaching future null infinity $\mathcal{J}^+$. Consequently, $g^{\hat u \hat u}=0$ is a necessary condition be satisfied by the coordinates. 
It is easy to see that this is not the case for the coordinates used  \eqref{eq:boosted_general_Kerr_Soares}.  
An equivalent statement for the existence of such one-form $\ell_\mu$ is that  for defining a metric be of Bondi-Sachs type, it has to obey the conditions $g_{rr}=g_{r\theta}=g_{r\phi}=0$ \cite{BSScholar}, which are violated in \eqref{eq:boosted_general_Kerr_Soares} by the the presence of  term $g_{r\phi}$.
What the author wishes to say is that then the Kerr metric in its out-going Eddington-Finkelstein form is recovered. 

If the parameter $a=0$,  the metric \eqref{eq:kerr-metric} reduces to the Schwarzschild solution expressed in outgoing-null polar coordinates (Eddington-Finkelstein): 
\begin{equation}\label{eq:sch-metric}
\begin{aligned}
    ds^2 =&
    -\Big(1-\f{2m}{\tilde r}\Big)d\tilde u^2
    -2d\tilde ud\tilde r
    +\tilde{r}^2
    \Big(d\tilde\theta^2+\sin^2\tilde\theta d\tilde\phi^2\Big)
    ,
    \end{aligned}
\end{equation}
Hereafter, we concentrate on the presentation in \cite{soares1} since all of our arguments can be  extended to show the invalidity of \cite{soares2} for general ``boosts'' with using the proper adaptations . 

For large values of $\tilde r$ on hypersurfaces $\tilde u = const$
\eqref{eq:kerr-metric} takes the form
\begin{equation}\label{asymptKS}
\begin{aligned}
    ds^2 =&
    \Big(\tilde{r}^2+a^2\cos^2\tilde{\theta}\Big)
    \Big(d\tilde\theta^2+\sin^2\tilde\theta d\tilde\phi^2\Big)
    \\&
    -2\Big(d\tilde u +a\sin^2\tilde\theta d\tilde \phi\Big)
    \Big(d\tilde r-a\sin^2\tilde\theta d\tilde \phi\Big)\\
    &
    -\Big(d\tilde u+a\sin^2\tilde\theta d\tilde\phi\Big)^2 
    +\mathcal{O}\Big(\f{m}{\tilde r}\Big)
    \end{aligned}
\end{equation}
which is a flat metric as can be shown by calculating the (vanishing) components of the  Riemann tensor at leading order.

Next, we recall: given  the standard Minkowski metric $\eta_{ab} = \mathrm{diag}(-1,1,1,1)$ in Cartesian coordinates $\tilde x^\mu=(\tilde t,\tilde x^i)$, its coordinate representation for an inertial observer in outgoing null coordinates in a rest frame follows from  the coordinate transformation, {$\tilde t =  \tilde u+\tilde r$}, $\tilde r^2=\delta_{ij}\tilde x^i\tilde x^j$, $\tilde x^i = ( \tilde r\sin \tilde \theta\cos \tilde \phi, \tilde r\sin \tilde \theta\sin \tilde \phi,  \tilde r\cos \tilde \theta)$  and has the form 
\begin{equation}\label{Intertial}
    ds^2 = -d \tilde u^2
    -2 d\tilde u d \tilde r 
    +  \tilde r^2 \Big(d\tilde \theta^2+\sin^2\tilde\theta d\tilde\phi^2\Big),
\end{equation}
see e.g. \cite{BKSS,BKS} for a recent discussion regarding boosted black holes and inertial frames. 
Metric \eqref{Intertial} is the inertial metric  $\eta_{\mu\nu}$ in outgoing polar null coordinate.   
If a general metric in outgoing null coordinates approaches the particular form of \eqref{Intertial} at large distances from the source, it is said that the asymptotic observer is in a Bondi frame \cite{Bondi,Sachs, BSScholar}. 

It is obvious that the leading order term of \eqref{asymptKS} is certainly not such Minkowski metric for $a\neq 0$. 
That is, if $a\neq0$, the coordinates used in \eqref{asymptKS} do not correspond those of an inertial observer. 
However, setting $a=0$, i.e. considering a non-rotating Kerr black hole a.k.a. the Schwarzschild black hole, \eqref{asymptKS} corresponds to the metric of an asymptotic inertial metric in null coordinates. 
Hereafter, we start considering the procedure of \cite{soares1} assuming $a=0$ and show that even in this case the resulting boosted Schwarzschild metric is not properly boosted with respect to an asymptotic observer in the associated inertial Bondi coordinates.

The ``boosted" Schwarzschild metric of \cite{soares1} (equation (23) in \cite{soares1} with $a=0$) is 
\begin{equation}\label{eq:boostedmetricsoares_a=0}
\begin{aligned}
    ds^2 =&\f{{ r}^2 (d\theta^2+\sin^2\theta d\phi^2)}{(A+B\cos\theta)^2}
    -2d  u  d  r 
    -\Big(1-\f{2m}{ r}\Big)d u  ^2.
    \end{aligned}
    \end{equation}
The first thing to note is that this metric is easily obtained by a simple change of {\it only one} of the angular coordinates in \eqref{eq:sch-metric}. This is achieved by setting $a=0$ in \eqref{eq:kerr-metric} and performing the coordinate transformation
\begin{equation}\label{soares}
\begin{aligned}
     \tilde u=&u\;\;,\;
    \tilde r=r\;,\;
    \tilde{\phi}=\phi.\\
    \cos \tilde\theta=&\f{B+A\cos(\theta)}{A+B\cos(\theta)},
\end{aligned}{}
\end{equation}
where $A^2-B^2=1$. 
According to \cite{soares1}, the functions $A$ and $B$ relate to the boost velocity $\beta$ like $\beta=B/A$ and the rapidity parameter $\zeta$ like $A=\cosh\zeta$ and  $B=\sinh\zeta$.  
Moreover,  it is never mentioned in \cite{soares1}  that their ``boosted" Kerr metric in their equation (23) can be  easily obtained applying the {\it same} transformation \eqref{soares} to the Kerr metric \eqref{eq:kerr-metric}, which is reproduced here for completeness 
\begin{equation}\label{eq:kerr-metricsoares}
\begin{aligned}
    ds^2 &=
    \f{{r}^2+\Sigma^2}{(A+B\cos\theta)^2}(d\theta^2+\sin^2\theta d\phi^2)
    \\&
    -2\Big[d u+\f{a\sin^2\theta}{(A+B\cos\theta)^2} d\phi\Big]\Big[dr-\f{a\sin^2\theta  d\phi}{(A+B\cos\theta)^2}\Big]\\
    &-\Big(1-\f{2mr}{r^2+\Sigma^2}\Big)\Big(du+\f{a\sin^2\theta  d\phi}{(A+B\cos\theta)^2}\Big)^2;
    \end{aligned}
    \end{equation}
where $\Sigma=a(B+A\cos\theta)(A+B\cos\theta)^{-1}$. 
In other words,
despite the claim of \cite{soares1}
that the `boosted' Kerr metric \eqref{eq:kerr-metricsoares} is obtained as an  {\it  exact stationary analytic solution}, we remark that it is just the original Kerr metric in different angular coordinates. 
We further note and demonstrate below that \eqref{soares}  is not a proper asymptotic Lorentz transformation, since a Lorentz transformation  does not only change the angular coordinates, but also the temporal and radial coordinates. 
In particular, the asymptotic Lorentz transformation maps one asymptotic inertial metric $\eta_{\mu\nu}(\tilde x^\alpha)$ to another asymptotic inertial metric  
$\eta_{\mu\nu}( x^\alpha)$.

It means that for large values of $r$, any asymptotically flat metric in Bondi coordinates  $\{u,r,\theta,\phi\}$  transforms to $\{\tilde u,\tilde r,\tilde \theta,\tilde \phi\}$  under the BMS group (an in particular under a Lorentz subgroup) like
\begin{equation}\label{eq:MappingInertialObservers}
\begin{aligned}
        &
        -d\tilde u ^2
        +2d\tilde u d\tilde r
        +\tilde r^2(d\tilde\theta^2+\sin^2\tilde\theta d\tilde\phi^2)+\mathcal{O}(1/\tilde r)\\
    =& -du^2
        -2dvd r
        + r^2(d\theta^2+\sin^2\tilde\theta d\phi^2)+\mathcal{O}(1/r).
    \end{aligned}
\end{equation}

For simplicity,  consider a boost in $z$ direction at large distances. Let $\{\tilde t,\tilde x,\tilde y,\tilde z\}$ the un-boosted Cartesian coordinates,  $\{t,x,y,z\}$  the boosted Cartesian coordinates and $(r, \theta, \phi)$ be the associated spherical coordinates in the boosted system with $r^2=\delta_{ij} x^ix^j$ and  $ x^i = (r\sin \theta\cos \phi,  r\sin \theta\sin \phi,  r\cos\theta)$).

Taking $\tilde t^\mu \partial_\mu = \partial_{\tilde t}$ as tangent vector to the world lines of the un-boosted observers,  corresponding boosted observers are tangent to $v^\mu = \gamma(1, \beta^i)$ with $\gamma =-v^\mu \tilde{t}_\mu= (1-\delta_{ij}\beta^i\beta^j)^{-1/2}$, so that  
the  Lorentz transformation for the coordinates $ x^\mu\rightarrow \tilde x^a$ and the radial functions $ r\rightarrow \tilde r$ are given  by \cite{fahnline}
 \begin{eqnarray}
  r^2 &\rightarrow \tilde r^2& =  x^\alpha  x_\alpha+(v^\alpha  x_\alpha)^2\\
  x^\mu &\rightarrow 
 \tilde x^\mu& = 
  x^\mu+ \frac{[t^\nu  x_\nu -(2\gamma +1)v^\nu  x_\nu]t^\mu }{1+\gamma}
\nonumber \\
 &&\;\;+\f{[t^\nu  x_\nu +v^\nu  x_\nu]v^\mu}{1+\gamma}\;\;.
 \end{eqnarray}
 For a (inverse)  boost in $z-$direction with $\beta^x=\beta^y=0$ and $\beta^z=\beta$, we find the relations 
  \begin{eqnarray}
  &&     \tilde u
      =\gamma\bigg[u+r(1+\beta\cos\theta)\bigg]\nonumber\\
      &&\;\;
      -r\sqrt{1+\gamma^2\bigg(\f{u}{r}+1+\beta\cos\theta\bigg)^2 -\Big(\f{u}{r}+1\Big)^2}
      \label{eq:wboosted_new}\\
  &&\tilde r
      =
      r\sqrt{1+\gamma^2\Big(\f{u}{r}+1+\beta\cos\theta\Big)^2 -\Big(\f{u}{r}+1\Big)^2}
      \label{eq:rboosted_new}\\
&&\cos(\tilde\theta)=
      \f{\tilde z}{\tilde r}\nonumber\\
      &=&\f{\gamma[\cos\theta+\beta(\f{u}{r}+1)]}{\sqrt{1+\gamma^2\Big(\f{u}{r}+1+\beta\cos\theta\Big)^2} -\Big(\f{u}{r}+1\Big)^2},\label{eq:thetaboosted_new}\\
     && \tilde \phi = \arctan\Big(\f{\tilde y}{\tilde x}\Big)=\arctan\Big(\f{ y}{ x}\Big) = \phi,\label{eq:phi_boosted_new}
  \end{eqnarray}
  between the un-boosted and boosted versions of the null coordinates $(\tilde u=\tilde{t}-\tilde{r}, \tilde r, \tilde \theta, \tilde \phi)\rightarrow(u=t-r,r,\theta,\phi)$. 
    For large distances (keeping $u$, $\theta$ and $\phi)$ fixed) \eqref{eq:wboosted_new} - \eqref{eq:phi_boosted_new} reduce to
    \begin{eqnarray}
    \tilde u&=&\f{u}{K(\theta)}+\mathcal{O}\Big(\f{1}{r}\Big);\;,\;\;\tilde r={K(\theta)}{r}+\mathcal{O}({r^0}),\label{eq:vr_asympt}\\
    \cos(\tilde\theta)&=&\f{\beta+\cos(\theta)}{1+\beta\cos(\theta)}+\mathcal{O}\Big(\f{1}{r}\Big)\;\;,\;\; \tilde \phi =\phi,\label{eq:abber}
    \end{eqnarray}
with $K(\theta)=\gamma(1+\beta\cos\theta)$.
Note, that the first part of \eqref{eq:abber} is the commonly known relativistic aberration formula. 
Relations \eqref{eq:vr_asympt} and \eqref{eq:abber} are the asymptotic Lorentz transformation for a boost along the $z-$axis. 
This transformation is a subset of a larger transformation,  which conform the BMS group. 
In fact, the BMS group is obtained in a more general framework by requiring a correspondent asymptotic behavior of the metric components when they are expressed in a Bondi system \cite{Bondi,Sachs,BSScholar}, and also in a geometrical way (see for example \cite{Moreschi_1986-AM}).

It is not difficult to check that this Lorentz transformation applied to \eqref{eq:kerr-metric} with $a=0$ maps the metric of an  asymptotic  inertial observer in coordinates $\tilde x^\mu $ to the metric of an  asymptotic  inertial observer in coordinates $ x^\mu $, as required by \eqref{eq:MappingInertialObservers}\footnote{To check this map, expressions for the $\mathcal{O}(r^{-1},r^0)$ terms in \eqref{eq:vr_asympt} and \eqref{eq:abber} are also needed. 
They can be easily found from \eqref{eq:wboosted_new}-\eqref{eq:thetaboosted_new}.}.
  The main point, we stress here, is  that  to make a Lorentz boost, a transformation in the $\tilde u$ and $\tilde{r}$ coordinates is needed. 
  However, Eqs.(5) do not contain this part of the Lorentz transformation. 
  Therefore, despite the claims of \cite{soares1}, the metric presented in that reference is not a properly boosted Kerr metric  with respect to the adapted coordinates of an asymptotic inertial frame, since the needed transformations are not even completely carried-out  in the Schwarzschild limit. 
  More generally, discarding supertranslations, BMS transformations in a neighborhood of null infinity can be written in terms of stereographic coordinates (whose relation to the standard spherical coordinates is $\zeta=e^{i\phi}\cot(\f{\theta}{2})$) as\footnote{In fact, the BMS group is defined {\it at} null infinity and is given only for the part of the transformation for the null coordinate and the angular coordinates charting null infinity. The  exact transformation of the radial coordinate depends of the kind of radial coordinate, which may be e.g. an area distance coordinate or an affine parameter.
  }
   \begin{eqnarray}
    \tilde u&=&\f{u}{K(\zeta,\bar\zeta)}+\mathcal{O}\Big(\f{1}{r}\Big);\;,\;\;\tilde r=K(\zeta,\bar\zeta){r}+\mathcal{O}({r^0}),\label{eq:vr_asympt2}\\
    \tilde\zeta&=&\f{a\zeta+b}{c\zeta+d}
    +\mathcal{O}\Big(\f{1}{r}\Big),\label{eq:abber2}
    \end{eqnarray}
  where $\{a,b,c,d\}$ are four complex parameters subject to the constraint $ac-bd=1$
  and
  $K(\zeta,\bar\zeta)$ is given by \cite{book:penrose-spinors-v1}
  \begin{equation}\label{eq:K_general_boost}
      K(\zeta,\bar\zeta)=\f{(a\zeta+c)(\bar a\bar\zeta+\bar c)+(b\zeta+d)(\bar b\bar\zeta+\bar d)}{1+\zeta\bar\zeta}.
  \end{equation}
  We remark that the ``generally boosted" Kerr metric presented in \cite{soares2}  can also be obtained from the Kerr metric \eqref{eq:kerr-metric} via the particular angular transformation \eqref{eq:abber2} associated to a general boost. However, as mentioned above, even in that situation this transformation is not sufficient to express the metric in a Bondi system. 
  Extra transformations  are necessary, because for a Bondi system $u$  must be a null surface forming coordinate, i.e. $u=\text{const}$ should define surfaces generated by  null vector fields reaching  $\mathcal{J}^+$. This is not the case for the $u$ coordinate present in \cite{soares1,soares2}. 
  
 In the Schwarzschild case of  \eqref{eq:boostedmetricsoares_a=0}, the $ u=\text{constant}$ hypersurfaces are indeed null surfaces reaching  null infinity.
  Nonetheless, the coordinates are not realizing a Bondi coordinate system either. In fact,  \eqref{eq:boostedmetricsoares_a=0} is expressed in a so called Newman-Unti coordinates (NU)\cite{Newman-Unti}. 
  More precisely, in terms of stereographic angular coordinates  the metric is a particular case of a more general family of metrics known as Robinson-Trautman geometries given by\cite{Dain:1996phot-rocket}
  \begin{equation}\label{eq:boostedmetricsoares_anueva=0}
\begin{aligned}
    ds^2 =&{ r}^2 \f{d\zeta d\bar\zeta}{(P_0 V)^2}
    -2d  u  d  r 
    -\Big(1-\f{2m}{ r}+\f{{V}_{,u}}{V}r\Big)d u  ^2,
    \end{aligned}
    \end{equation}
     with $P_0=1+\zeta\bar\zeta$, $V=V(u,\zeta,\bar\zeta)$ and $m=m(u)$. 
    These metrics belong to the class of Robinson-Trautman solutions defined  by the property that they  admit a geodesic, shear-free and twist-free but expanding null congruence.
     Regarding \eqref{eq:boostedmetricsoares_a=0}, we have $m_{,u}=0$ and 
     \begin{equation}\label{eq:V_soares}
         V = A+B\cos\theta 
         = A+ B \f{\zeta\bar\zeta-1}{1+\zeta\bar\zeta},
     \end{equation}
     showing that also $V_{,u}=0$. 
     Moreover, the coordinates $\{u,r,\zeta,\bar\zeta\}$  correspond to a Bondi system only if $V=1$ (rest frame).
     Note, we are not saying that the metric \eqref{eq:boostedmetricsoares_anueva=0} could not be interpreted as a boosted black hole; what we are saying is that if these NU coordinates are used we must yet to relate it to a Bondi system in order to extract physical quantities. 
     For example,  
     as discussed in \cite{momentum}, the total linear momentum $P^\alpha$ for the metric \eqref{eq:boostedmetricsoares_anueva=0} can be computed in a non-Bondi system from the formula
     \begin{equation}
     P^\alpha=\int \f{m}{V^3}\hat{\ell}^a dS^2 
     \end{equation} 
 with $dS^2$ the surface element of a unit sphere and
 \begin{equation}
     \hat\ell^a=\Big(1,\f{\zeta+\bar\zeta}{1+\zeta\bar\zeta},\f{\zeta-\bar\zeta}{i(1+\zeta\bar\zeta)},\f{\zeta\bar\zeta-1}{1+\zeta\bar\zeta}\Big).
 \end{equation} 
 Note that this expression was also correctly used in \cite{soares2} to compute the four-momentum of its metrics.
 
 However, some of the analysis carried out on the metrics \cite{soares1,soares2} is misleading.
For example, the location of the horizon for the `boosted' metric \eqref{eq:kerr-metricsoares} is  measured to take the same value as in the Kerr metric.
 This was interpreted as being a consequence that a boost does not change null surfaces.
 It is  true that boosts do not distort null surfaces, but its coordinate representation  for an asymptotic boosted inertial observer, however, would be in general different.
 The  reason why the coordinate location of the horizon for the `boosted' Kerr metric \eqref{eq:kerr-metricsoares} takes the same value as in the Kerr metric is because the radial coordinate was not changed by the coordinate transformation (c.f. \eqref{soares}).
 Notwithstanding, it is well-known that
 the  shapes of the boosted vs. unboosted horizon  is coordinate dependent (see. e.g \cite{PhysRevD.66.084024,Akcay:2007vy}).
  We note, if we were to attempt a similar procedure as in \cite{soares1,soares2} for the location of a photon sphere $S_\text{ph}$ in the boosted Schwarzschild metric \eqref{eq:boostedmetricsoares_a=0}, we  would find it placed at the same radial coordinate $r=3m$ as in the un-boosted black hole, even when for this case the surface $S_\text{ph}$ is not a null hypersurface.\footnote{The same could be said for other special orbits, as for example the inner stable circular orbit (ISCO) of a test massive particle.} Again, it is only because of we are not properly transforming the radial and timelike coordinates.
 
 In \cite{soares1}, it  is  claimed that {\it ``The boosted Kerr geometry also presents an ergosphere,...''};
 this is not  surprising at all because  \cite{soares1}'s metric is the Kerr metric after the coordinate transformation \eqref{soares}. 
  The coordinate expression for the  ergosphere  of \cite{soares1,soares2}  shows a most complex dependence from the angular coordinates.
Again, the relevant expression is  analyzed by  using  the un-boosted (Kerr) radial coordinate  and the  `boosted' angular coordinates. 
That is,  there is again no proper use of the associated `boosted' radial coordinate. 

In any case, the geometrical definition of the ergosphere of the Kerr black hole is given by the set of points, where the (global) timelike Killing vector $\f{\partial}{\partial u}$ becomes a null vector.
This is a geometrical (coordinate-independent) definition. 
However, it is clear that for the analysis of ergosphere of a boosted Kerr black hole by an asymptotic observer,  associated inertial coordinates $\{u,r,\theta,\phi\}$ should be used instead of the  mixed set of coordinates $\{\tilde{u},\tilde{r},\theta,\phi\}$ like in \cite{soares1,soares2}.\footnote{It is worthwhile to emphasize that for an asymptotic observer  there exists another notion of (observer dependent) ergosphere based on the asymptotic Killing vector aligned with the asymptotic observer, which again should be expressed in adapted coordinates of this observer (See for example \cite{Penna:2015qta} where an analysis of these `resulting  ergospheres' of boosted Schwarzschild black hole can be found). Let us note that for the metric \eqref{eq:boostedmetricsoares_a=0} these kind of ergospheres of \cite{Penna:2015qta}  can not be obtained from the procedure followed by \cite{soares1,soares2}.}

  We  also stress the well known fact that Kerr's original metric does not approach the Minkowski metric of an inertial observer for large radii (also seen in \eqref{asymptKS}). 
  Hence, it ought not be used for the discussion of physical effects resulting from a comparison of boosted and un-boosted black holes in the asymptotic regime.
  In fact, to unambiguously define a boost, an inertial observer needs to be able to singled out, so that it is clear with respect to which rest frame the boost is performed. 
  Henceforth, one wishes to cast the black hole metric $g_{\mu\nu}$ to be boosted  into a form like 
     $ g_{\mu\nu} = \eta_{\mu\nu}+ \hat g_{\mu\nu}$
  where $\eta_{\mu\nu}$ is the Minkowski metric and $\hat g_{\mu\nu}$ is a function of the coordinates. 
  Such a representation of  $g_{\mu\nu}$ can be obtained two different ways: (i) a linearization and (ii) finding a Kerr-Schild representation of the black hole metric. 
 The linearisation (i) covers three branches. 
 One realisation of (i) is the introduction of a ``smallness''
 parameter $\epsilon$ measuring the deviation from flat spacetime (i.e. $0\approx\epsilon\approx | \hat g_{\mu\nu}|$ for every component of $\hat g_{\mu\nu}$). 
 The second realisation is the assumption that at given distance from the black hole an inertial observer is introduced and Fermi normal coordinates \cite{Manasse} are constructed around the worldline of the observer.
 While the third realization is assuming that in a given hypersurface of the corresponding spacetime there is  a (radial) function $r$ constructed from the local coordinates.
 This coordinate should have the property that for large values of this function the metric approaches a Minkowski metric (i.e. $\hat g_{\mu\nu}$ becomes small for $r\rightarrow\infty$) 
 In fact, the Boyer-Lindquist form as well as the Kerr-Schild form of the Kerr metric have this property for large values of $r$. 
 
 On top of that, Kerr-Schild metric have the property that the metric is written as
 $g_{\mu\nu} = \eta_{\mu\nu}+Hk_\mu k_\nu$, where $H$ is a scalar function and  $k_\mu$ is a null vector with respect to $\eta_{\mu\nu}$ and $g_{\mu\nu}$. 
  Such ansatz was, in fact, first used by Trautman in the study of radiative spacetimes \cite{Trautman} and it was crucial for finding of the Kerr solution \cite{KerrDiscovery}.     
 In particular, it had recently been pointed out that the spacetimes of the Schwarzschild and  Kerr black hole in Kerr-Schild form have not only one  inertial frame serving as a background spacetime to define a boost, but two\footnote{An brief remark about this fact can already be found the in Boyer-Lindquist paper \cite{boyer}.} \cite{BKSS,BKS}. 
  These two Minkowski backgrounds are tied to the outgoing and outgoing principal null directions of the respective metric in Kerr-Schild form. 
  The inertial coordinates of these Minkowski backgrounds transform between each other via a non-linear coordinate transformation. 
  Indeed it was shown in \cite{ST_RM&AM,BKSS,BKS} that for the correct value of the boost memory at {\it future} null infinity, the discussion of the boost must be done in the Minkowski background of the {\it ingoing} formulation. 
 
  For a Schwarzschild/Kerr black hole which is initially at rest and then ejected with mass $m$ and velocity $\beta$ along the $z-$axis, the boost memory at null infinity is \cite{thorne,ST_RM&AM,BKSS,BKS} \label{eq:supertranslation}
  \begin{equation}\label{eq:memory}
     \Delta  \sigma  = \frac{4\gamma m \beta^2\sin^2 \theta}{1-\beta\cos\theta}.
  \end{equation}
 The supertranslation $\alpha$ relating the retarded time cuts cuts $u=\infty$ and $u=-\infty$ at null infinity  is \cite{ST_RM&AM}
 \begin{equation}\label{eq:supertranslation}
     \alpha = 4m\gamma(1-\beta\cos\theta)\ln(1-\beta\cos\theta)
 \end{equation}
  Above relations \eqref{eq:memory} and \eqref{eq:supertranslation} can by no means be reproduced from expression \eqref{eq:kerr-metricsoares}. 
  
 \section{Acknowledgments} 
 EG thanks CONICET and Secyt-UNC for financial support. TM thanks FaMaF-UNC, C\'ordoba for hospitality and the University Diego Portales for a travel grand.
 TM also appreciates P. Jofr\'e for support.
  
\bibliographystyle{unsrt}
 \bibliography{bib} 
 \end{document}